\documentclass[a4paper]{jpconf}
\usepackage{graphicx}

\def\be{\begin{equation}}
\def\ee{\end{equation}}

\begin{document}
\title{Coherent search of continuous gravitational wave signals: extension of the 5-vectors method to a network of detectors}

\author{P Astone$^1$, A Colla$^{2,1}$, S D'Antonio$^3$, S Frasca$^{2,1}$, C Palomba$^1$}

\address{$^1$ INFN, Sezione di Roma, I-00185 Roma, Italy}
\address{$^2$ Universita' di Roma "Sapienza", I-00185 Roma, Italy}
\address{$^3$ INFN, Sezione di Roma 2, I-00133 Roma, Italy}

\ead{cristiano.palomba@roma1.infn.it}

\begin{abstract}
We describe the extension to multiple datasets of a coherent method for the search of continuous gravitational wave signals, based on the computation of 5-vectors. In particular, we show how to coherently combine different datasets belonging to the same detector or to different detectors. In the latter case the coherent combination is the way to have the maximum increase in signal-to-noise ratio. If the datasets belong to the same detector the advantage comes mainly from the properties of a quantity called {\it coherence} which is helpful (in both cases, in fact) in rejecting false candidates. The method has been tested searching for simulated signals injected in Gaussian noise and the results of the simulations are discussed.
\end{abstract}

\section{Introduction}
The search for continuous gravitational wave signals, like those emitted by a deformed rotating neutron star, is an important target of the analysis of data produced by interferometric detectors, like GEO \cite{geo600}, LIGO \cite{ligo} and Virgo \cite{virgo}. In the so-called {\it targeted search} the source parameters (sky position, frequency and frequency derivatives) are assumed to be known. This is the case, .e.g., of the search of signals from known pulsars for which radio-astronomical observations can provide very accurate measures of the parameters themselves. For this kind of search coherent data analysis methods, which precisely correct the signal phase variations due to Doppler, spin-down, Einstein and, if necessary, Shapiro effects, allow in principle to approach the best sensitivity at a relatively small computing cost. Various coherent pipelines have been developed and have been used (and will be used) to analyze data from Virgo and LIGO scientific runs, see e.g. \cite{vela_pap}. In particular, we consider here a method which goes through the computation of the so-called 5-vectors \cite{asto1} which has been used in the analysis of Virgo VSR2 data for the search of continuous gravitational wave signals from the Vela pulsar \cite{vela_pap}. 

Given the expected weakness of continuous signals, it is important to use the available data, which are typically affected by various kinds of disturbances, in the most effective way. This is done in the 5-vectors method by applying different techniques. For instance, a cleaning procedure is used both on the initial time domain data \cite{asto2} and after Doppler, spin-down and Einstein effect corrections, in order to remove particularly disturbed periods. Moreover, a Wiener filter is applied for taking into account slow non-stationarities so to weight less the more noisy periods \cite{vela_pap}.  In this paper we describe one more analysis step that can be applied in order to improve the sensitivity of the search: if more datasets are available, belonging to the same detectors (e.g. two different runs) or to different detectors, they can be properly combined in order to increase the output signal-to-noise ratio and to improve rejection of false candidates. Other methods of coherently combining data from different detectors have been developed in the past, like e.g. the network analysis using the $\mathcal F$-statistic  \cite{cutler} and a complex heterodyne method \cite{dupuis}. 

The plan of the paper is the following. In Sec.(\ref{5vect}) the basic concepts of the 5-vectors method are summarized and the definition of signal-to-noise ratio is given. In Sec.(\ref{5nvect}) the coherent combination of more datasets is introduced and tests done with simulated data are described. Sec.(\ref{conc}) is devoted to discuss further future developments of the method.

\section{The 5-vectors method}
\label{5vect}
Here we briefly recall the 5-vectors method and define the signal-to-noise ratio (SNR). More details can be found in \cite{asto1} and \cite{vela_pap}.
Let us indicate the data to be analyzed with 
\begin{equation}
x'(t)=n'(t)+h'(t)
\end{equation}
where $n'(t)$ is the noise and $h'(t)$ is a continuous gravitational wave signal. The data, after the extraction of a small band around the expected frequency of the signal, are corrected for Doppler, spin-down and Einstein effects by a re-sampling procedure described in detail in \cite{vela_pap} and which effectiveness is demonstrated by the very good recovery of parameters of hardware injected pulsar signals in Virgo VSR2 data as discussed in \cite{vela_pap}. Let us indicate the corrected data with $x(t)=n(t)+h(t)$.  At this stage a signal present into the data is monochromatic apart from an amplitude and phase sidereal modulation. The gravitational wave strain at the detector can be then written as
\begin{equation}
h(t)=H_0(H_+A^+(t)+H_{\times}A^{\times}(t))e^{j(\omega_0t+\Phi_0)}
\label{hoft}
\end{equation}
(where the operation of taking the real part is understood in order to have the physical strain)
In this equation the two amplitudes are given by
\begin{eqnarray}
H_+ &=& \frac{\cos{(2\psi)}-j\eta~ \sin{(2\psi)}}{\sqrt{1+\eta^2}} \nonumber \\
H_{\times} &=& \frac{\sin{(2\psi)}+j\eta~ \cos{(2\psi)}}{\sqrt{1+\eta^2}} 
\end{eqnarray}
where $\eta$ is the ratio between the minor and the major semi-axis of the wave polarization ellipse ($\eta=0$ for a linearly polarized signal, $\eta\pm 1$ for a circularly polarized signal) and $\psi$ is the polarization angle defining the direction of the ellipse semi-major axis respect to the celestial parallel of the source (measured counterclockwise). The functions $A^+(t),~A^{\times}(t)$ are equal to the standard beam-pattern functions $F^+(t;\psi),~F^{\times}(t;\psi)$, defined e.g. in \cite{jks}, computed at $\psi=0$. The power of the signal in Eq.(\ref{hoft}) is spread among the five angular frequencies $\omega_0,~\omega_0\pm \Omega,~\omega_0\pm 2\Omega$, where $\Omega$ is the Earth sidereal angular frequency.

The noise $n(t)$ is assumed to be gaussian with mean $\mu=0$ and standard deviation $\sigma$. The signal detection and parameter estimation procedure is based on the computation of data and signal 5-vectors, defined as their Fourier components at the 5 frequencies in which the signal is split by the sidereal modulation. More precisely, given a generic time series $f(t)$, the corresponding 5-vector is a complex vector
\be
\mathbf{F}=\int_T f(t)e^{-j(\omega_0-\mathbf{k}\Omega)t}dt
\label{def5vect}
\ee
where $\mathbf{k}=[-2,-1,...,2]$ and $T$ is the observation time. 
The signal 5-vector is defined as
\be
\mathbf{A}=\left({H}_{+}\mathbf{A}^{+}+{H}_{\times}\mathbf{A}^{\times}\right)
\label{signal5vect}
\ee
Note that for a source with known position and frequency and for a given detector $\mathbf{A}^{+}$ and $\mathbf{A}^{\times}$ can be computed  from Eq.(\ref{def5vect}) while $H_0,~\eta,~\psi,~\Phi_0$ are in general unknown (for a few pulsars, like Vela and Crab, astronomical observations provide estimations of $\eta$ and $\psi$ \cite{ng}). The data 5-vector can be expressed as
\be
\mathbf{X}=H_0e^{j\Phi_0}\left({H}_{+}\mathbf{A}^{+}+{H}_{\times}\mathbf{A}^{\times}\right)+\mathbf{N}=H_0e^{j\Phi_0}\mathbf{A}+\mathbf{N}
\ee
where $\mathbf{N}$ is the noise 5-vector. It is important to stress that data or signal 5-vectors contains exactly the same information of the corresponding time series.
Once the 5-vectors of data, $\mathbf{X}$, and signal, $\mathbf{A}^{+}$ and $\mathbf{A}^{\times}$, have been computed the following complex amplitude estimators are built:
\begin{eqnarray}
\hat{H}_{+}=\frac{\mathbf{X}\cdot \mathbf{A}^{+}}{|\mathbf{A}^{+}|^2} \nonumber \\
\hat{H}_{\times}=\frac{\mathbf{X}\cdot \mathbf{A}^{\times}}{|\mathbf{A}^{\times}|^2}
\label{hphc}
\end{eqnarray}
where the scalar product among two complex vectors $\mathbf{a}$ and $\mathbf{b}$, each with $m$ components, is defined as $\mathbf{a}\cdot \mathbf{b}=\sum_{i=1}^m a_ib_i^*$.
$\hat{H}_+$ and $\hat{H}_{\times}$ are the estimation of, respectively, $H_0e^{j\Phi_0}H_+$, and $H_0e^{j\Phi_0}H_{\times}$ and are used to compute the detection statistics
\be
S=\left|\mathbf{A}^{+}\right|^4\left|\hat{H}_{+}\right|^2+\left|\mathbf{A}^{\times}\right|^4\left|\hat{H}_{\times}\right|^2
\label{detstat}
\ee
which is compared to its expected distribution in presence of noise alone to establish if a statistically significant candidate is present. In case of detection Eq.(\ref{hphc}) is also used to estimate source parameters.  In particular, the estimation of the signal amplitude is given by
\be
\hat{H}_0=\sqrt{|\hat{H}_{+}|^2+|\hat{H}_{\times}|^2}
\label{H0est}
\ee
A preliminary comparison of our detection statistic with the $\mathcal F$-statistic is done in \cite{asto1}. A more detailed comparison, also with other detection statistics like the $\mathcal B$-statistic introduced in \cite{prix}, will be done in a forthcoming paper.

If a detection is claimed we can quantify its reliability by a parameter called {\it coherence} defined as \cite{asto1}
\be
c=\frac{\hat{H}\mathbf{\hat{A}}}{|\mathbf{X}|^2}
\label{cohe}
\ee
where $\mathbf{\hat{A}}=\hat{H}_{+}\mathbf{A}^{+}+\hat{H}_{\times}\mathbf{A}^{\times}$ is the estimated total signal 5-vector and 
$\hat{H}=\frac{\mathbf{X}\cdot \mathbf{\hat{A}}}{|\mathbf{\hat{A}}|^2}$. The {\it coherence} is a number between 0 and 1 which measures the resemblance between the shape of the expected signal and the data. In fact it does not depend on scaling factors on the signal but just on its shape. We have numerically found that in the absence of any signal the {\it coherence} is described (assuming all the source polarization parameters are unknown) by a $\beta$ distribution with parameters $\alpha_1=2,~\alpha_2=3$:
\be
p(c;\alpha_1=2,\alpha_2=3) =\frac{12c\cdot (1-c)^2}{B(\alpha_1=2,\alpha_2=3)}
\label{cohe1}
\ee
where $B(\alpha_1,\alpha_2)$ is the $\beta$ function which acts as a normalization constant. Large values of {\it coherence} have a small probability of being produced by noise alone.
We will further understand the utility of the {\it coherence} when extending the analysis to more datasets.

If no significant candidate is found an upper limit on the signal amplitude can be computed through a Monte Carlo simulation \cite{vela_pap}. 

The 5-vectors method presents various strong points. For instance, the computation of cross products among data and signal templates is very fast because it involves only 10 complex numbers instead of $o(10^7)$ as it would be working in the time domain and this, in conjunction with the re-sampling procedure used to correct the Doppler effect, is particularly useful when doing simulations. Moreover, the re-sampling procedure allows to easily extend the method to the so-called 'narrow band' searches in which a possible mismatch between the electromagnetic and gravitational wave signal can be accounted for. As we will see in Sec.\ref{5nvect}, another advantage is the
straightforward extension to a network of detectors, which is the main topic of this paper.
  
\subsection{Signal-to-noise ratio}
In the absence of signals, assuming the noise has a Gaussian distribution with mean zero and variance $\sigma^2$, each component $\mathbf{X}_i~(i=1,..5)$ of the data (noise) 5-vector is distributed as a complex gaussian variable, with $\mu_X=0$ and $\sigma^2_{X}=\sigma^2T$, as it can be immediately seen from Eq.(\ref{def5vect})  (i.e. the real and imaginary parts are two independent gaussian variables with mean zero and variance $\sigma^2T/2$). The expectation value of $|\mathbf{X}|^2$ is $E[|\mathbf{X}|^2]=\sum_{i=1}^5|\mathbf{X}_i|^2=5\sigma^2T$. Using Eq.(\ref{hphc}) it is easy to see that the two amplitude estimators also have a gaussian distribution with mean zero and variance 
\begin{eqnarray}
\sigma_{+}^2 &=& E\left[\left|\hat{H}_{+}\right|^2\right]=\frac{\sigma^2T}{|\mathbf{A}^{+}|^2}\nonumber \\
\sigma_{\times}^2 &=& E\left[\left|\hat{H}_{\times}\right|^2\right]=\frac{\sigma^2T}{|\mathbf{A}^{\times}|^2}
\label{varfiltcomp}
\end{eqnarray}
From Eq.(\ref{H0est})
\be
\sigma^2_{\hat{H}_0}=E\left[\left|\hat{H}_{0}\right|^2\right]=E\left[\left|\hat{H}_{+}\right|^2\right]+E\left[\left|\hat{H}_{\times}\right|^2\right]=
\sigma^2_++\sigma^2_{\times}=\sigma^2T\frac{|\mathbf{A}^{+}|^2+|\mathbf{A}^{\times}|^2}{|\mathbf{A}^{+}|^2|\mathbf{A}^{\times}|^2}
\label{vartot}
\ee
Let us define the signal-to-noise ratio ($SNR$) at the filter output as the ratio between the estimated signal amplitude and the standard deviation of the estimation when only noise is present:
\be
SNR=\frac{\hat{H}_0}{\sigma_{\hat{H}_0}}=\frac{\hat{H}_0|\mathbf{A}^{+}||\mathbf{A}^{\times}|}{\sigma\sqrt{(|\mathbf{A}^{+}|^2+|\mathbf{A}^{\times}|^2)T}}
\label{snr}
\ee
From the definition of 5-vector, see Eq.(\ref{def5vect}), we see that each component of the signal 5-vectors $\mathbf{A}^{+}$ and $\mathbf{A}^{\times}$ increases linearly with the observation time $T$. Hence, from Eq.(\ref{snr}), we have $SNR\propto \sqrt{T}$, as expected for a coherent analysis. 

\section{Extension to more datasets}
\label{5nvect}
Let us consider $n$ datasets. Let us assume for simplicity that they have all the same length, the same noise level and belong to the same detector. Let us compute the corresponding 5-vectors, $\mathbf{X}_i,~i=1,...n$. Given a source target let us compute also the signal 5-vectors, $\mathbf{A}^+_i,~\mathbf{A}^{\times}_i,~i=1,...n$. We can now put together the $n$ 5-vectors of each kind in one 5$n$-vector:
\begin{eqnarray}
\mathbf{X} &=& \left[\mathbf{X}_1,\mathbf{X}_2,...\mathbf{X}_n\right] \nonumber \\ 
\mathbf{A}^{+} &=& \left[\mathbf{A}^{+}_1,\mathbf{A}^{+}_2,...\mathbf{A}^{+}_n\right] \nonumber \\
\mathbf{A}^{\times} &=& \left[\mathbf{A}^{\times}_1,\mathbf{A}^{\times}_2,...\mathbf{A}^{\times}_n\right]
\end{eqnarray}
In this simple case $|\mathbf{A}^{+}_i|=|\mathbf{A}^{+}_j|,~|\mathbf{A}^{\times}_i|=|\mathbf{A}^{\times}_j|~~\forall i,j=1,...n$ (because the datasets belong to the same detector and have the same length) so that 
\begin{eqnarray}
\left |\mathbf{A}^{+}\right|^2 &=& n\left |\mathbf{A}^{+}_i\right|^2~~~\forall i=1,...n \nonumber \\
\left |\mathbf{A}^{\times}\right|^2 &=& n\left |\mathbf{A}^{\times}_i\right|^2~~~\forall i=1,...n
\end{eqnarray}
Then from Eq.(\ref{snr}) we have  
\be
SNR=SNR_i\cdot \sqrt{n}~~~\forall i=1,...n
\ee
This means that working with 5n-vectors the signal-to-noise ratio increases as in the case of a coherent analysis over the total observation time.
This result can be easily generalized to the case in which datasets of different length, sensitivity or belonging to different detectors are considered.
For instance, let us consider two datasets with different sensitivities, i.e. with two different values of variance $\sigma^2_1,~\sigma^2_2$ for the components of the corresponding 5-vectors, and two different pairs of signal templates $(\mathbf{A}^+_1,~\mathbf{A}^{\times}_1)$ and $(\mathbf{A}^+_2,~\mathbf{A}^{\times}_2)$ (different because corresponding to different detectors and/or different observation periods). In this case the variance of the signal amplitude estimation is given by
\be
\sigma^2_{\hat{H}_0}=\frac{\sigma^2_1|\mathbf{A}^{+}_1|^2+\sigma^2_2|\mathbf{A}^{+}_2|^2}{\left(|\mathbf{A}^{+}_1|^2+|\mathbf{A}^{+}_2|^2\right)^2}+
\frac{\sigma^2_1|\mathbf{A}^{\times}_1|^2+\sigma^2_2|\mathbf{A}^{\times}_2|^2}{\left(|\mathbf{A}^{\times}_1|^2+|\mathbf{A}^{\times}_2|^2\right)^2}
\ee
It can be shown that also the probability distribution of the detection statistics is equal to that we would have in case of 5-vector analysis over the whole observation time. The estimators of the signal complex amplitudes are given by
\begin{eqnarray}
\hat{H}_{+} &=& \frac{\mathbf{X}\cdot \mathbf{A}^{+}}{|\mathbf{A}^{+}|^2}=\frac{\sum_{i=1}^n \mathbf{X}_i\cdot \mathbf{A}^{+}_i}{\sum_{i=1}^n |\mathbf{A}^{+}_i|^2} \nonumber \\
\hat{H}_{\times} &=& \frac{\mathbf{X}\cdot \mathbf{A}^{\times}}{|\mathbf{A}^{\times}|^2}=\frac{\sum_{i=1}^n \mathbf{X}_i\cdot \mathbf{A}^{\times}_i}{\sum_{i=1}^n |\mathbf{A}^{\times}_i|^2}
\end{eqnarray}
From these equations in case of detection we can estimate signal parameters exactly in the same way as in the $n=1$ case.
We also see that each data 5-vector 'interacts' only with the corresponding templates. 

\begin{figure}
\begin{tabular}{c}
\includegraphics[width=1.0\textwidth, height=90mm]{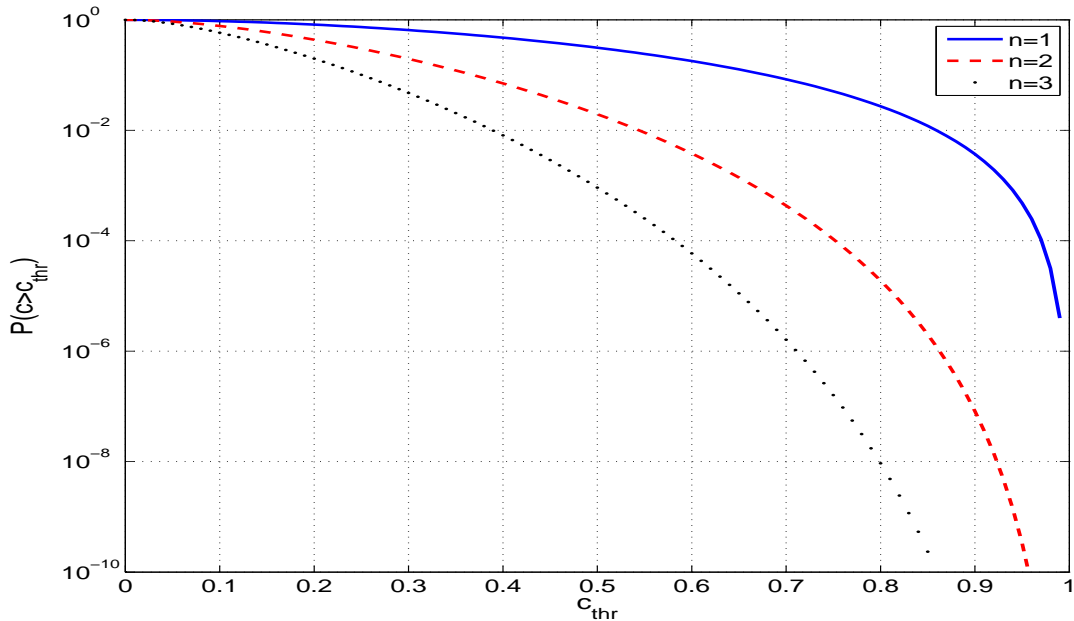} 
\end{tabular}
\caption{Probability to have a value of the {\it coherence} larger than a given threshold $c_{thr}$ for a 5n-vector analysis with different values $n$ of the number of datasets.}\label{cohe_cum_fig}
\end{figure}

The practical usefulness of working with 5n-vectors is clear when we want to combine data from different detectors, but it can be appreciated also when we have two or more datasets of the same detector. In this case the advantage of using 5n-vectors comes mainly from the properties of the {\it coherence}, defined by Eq.(\ref{cohe}). We have numerically found that the probability distribution of the {\it coherence} computed over 5n-vectors in presence of gaussian noise alone is
a $\beta$ distribution with parameters $\alpha_1=2,~\alpha_2=5n-2$:
\be 
p(c;\alpha_1=2,\alpha_2=5n-2)= \frac{(5n-2)(5n-1)c\cdot (1-c)^{5n-3}}{B(\alpha_1=2,\alpha_2=5n-2)}
\ee
which for $n=1$ reduces to Eq.(\ref{cohe1}) . The probability of obtaining a value of {\it coherence} larger than a given threshold $c_{thr}$ is 
\be
P(c>c_{thr})=(5n-1)(1-c_{thr})^{5n-2}-(5n-2)(1-c_{thr})^{5n-1}
\label{cohe_cum}
\ee
and then it is a decreasing function of the number $n$ of datasets being considered. This is shown in Fig.(\ref{cohe_cum_fig}) where Eq.(\ref{cohe_cum}) is plotted for $n=1,2,3$. For instance, the probability of getting a value of the {\it coherence} larger than $c_{thr}=0.8$ is $\sim 0.03$ for $n=1$, and decreases to $\sim 10^{-5}$ for $n=2$ and to $\sim 10^{-8}$ for $n=3$. The computation of the {\it coherence} after a 5n-vector analysis is then a powerful tool to reject false candidates produced by noise fluctuations. 

We have tested the 5n-vector method by analyzing data in which a simulated signal has been added to several realizations of white Gaussian noise. Tab.(\ref{tab1}) shows the mean value and variance of the distribution of estimated signal parameters for a set of 500 realizations of noise (with output $SNR\simeq 15$). Columns 1 and 2 refer to the 5-vector analysis of each of the two datasets, which have the same length and same sensitivity in this example. Column 3 refers to the 5n-vector analysis over the two periods (n=2 in this case) while column 4 refers to the 5-vector analysis over the total dataset. We see that the results of the 5n-vector method are identical to those obtained coherently analyzing the whole observation period. On the other hand, while also in the case of analysis over the single periods the mean value of estimated parameters is very near to the injected ones, due to the high SNR we have in this example, the variance of the estimations is larger. In fact, it decreases as the total observation time and so, as expected, for the 5n-vector analysis and for the coherent analysis over the total dataset it reduces by a factor of about 2.   

\begin{table}[h]
\caption{Parameter estimation in the analysis of simulated data. The simulation consisted in the injection of a simulated signal into 500 realizations of white Gaussian noise. Injections were done into two datasets of length 100 days assumed to belong to the Virgo detector. The noise level was the same (output $SNR\simeq 15$). The injected parameters were: $H_0=1,~\eta=0.16,~\psi=24.439^o$. In columns 1 and 2 the mean value and variance of the estimated parameters obtained with the 5-vect analysis applied separately to the two datasets are given. In column 3 the results for the 5n-vect analysis over the two periods are given (with n=2 in this case). Column 4 refers to the 5-vector analysis of the whole period covered by the two datasets.}
\centering
\begin{tabular}{c c c c c}
\hline\hline
 & dataset 1  & dataset 2 & 5n-vect & total\\ [0.5ex]
\hline
$\bar{H}_0$ & 1.0071 & 1.0138 & 1.0070 & 1.0070 \\
$\sigma^2_{H_0}$ & 0.0045 & 0.0045 & 0.0025 & 0.0025 \\
$\bar{\eta}$ & 0.1568 & 0.1601 & 0.1587 & 0.1587 \\
$\sigma^2_{\eta}$ & 0.0048 & 0.0047 & 0.0023 & 0.0023 \\
$\bar{\psi}$ & 25.4491 & 25.4888 & 25.4746 & 25.4746 \\ 
$\sigma^2_{\psi}$ & 4.3798 & 3.9501 & 2.0240 & 2.0240 \\ [1ex]
\hline
\end{tabular}
\label{tab1}
\end{table}

\begin{figure}
\begin{tabular}{cc}
\includegraphics[width=.5\textwidth, height=45mm]{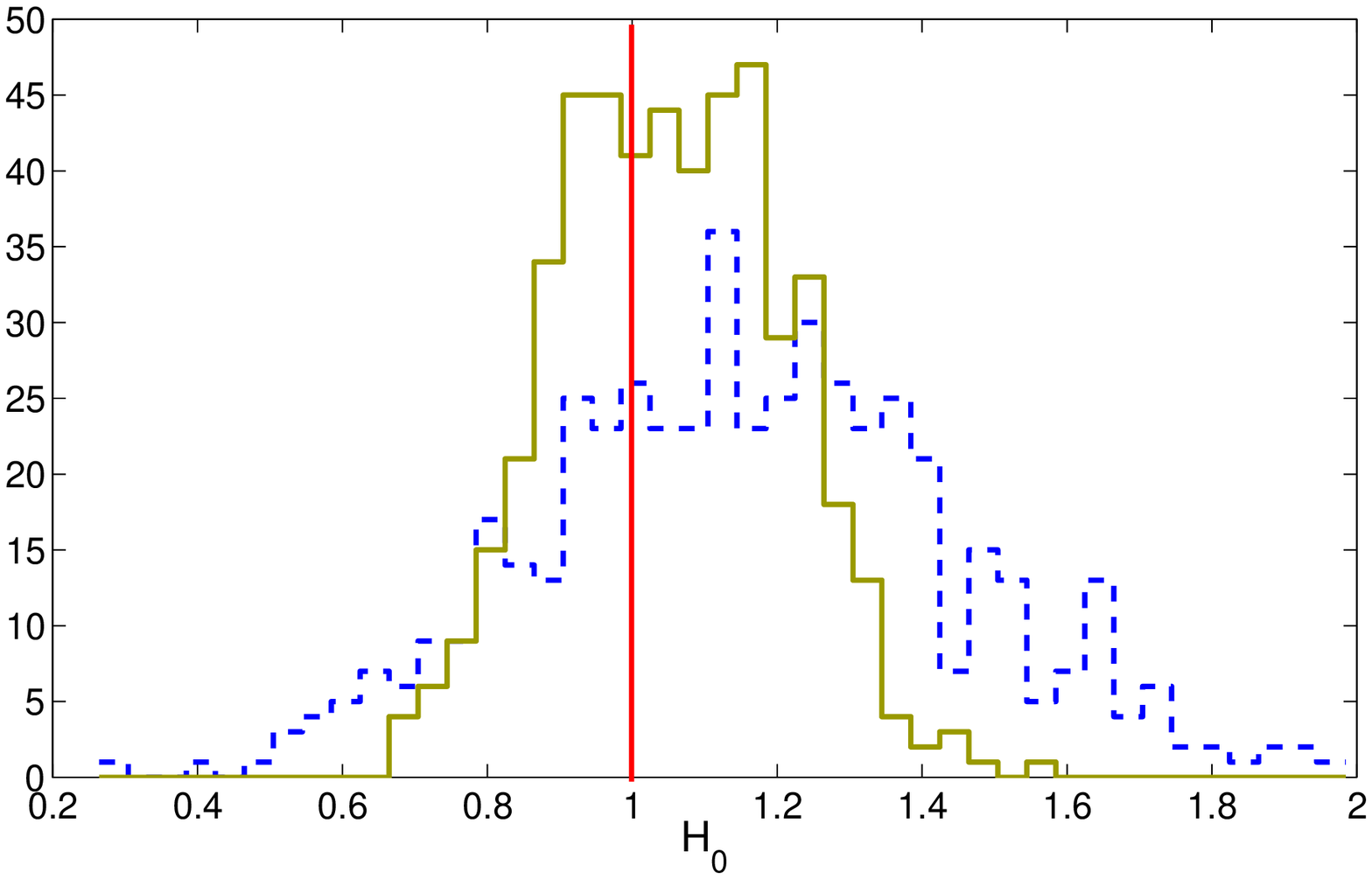} &
\includegraphics[width=.5\textwidth, height=45mm]{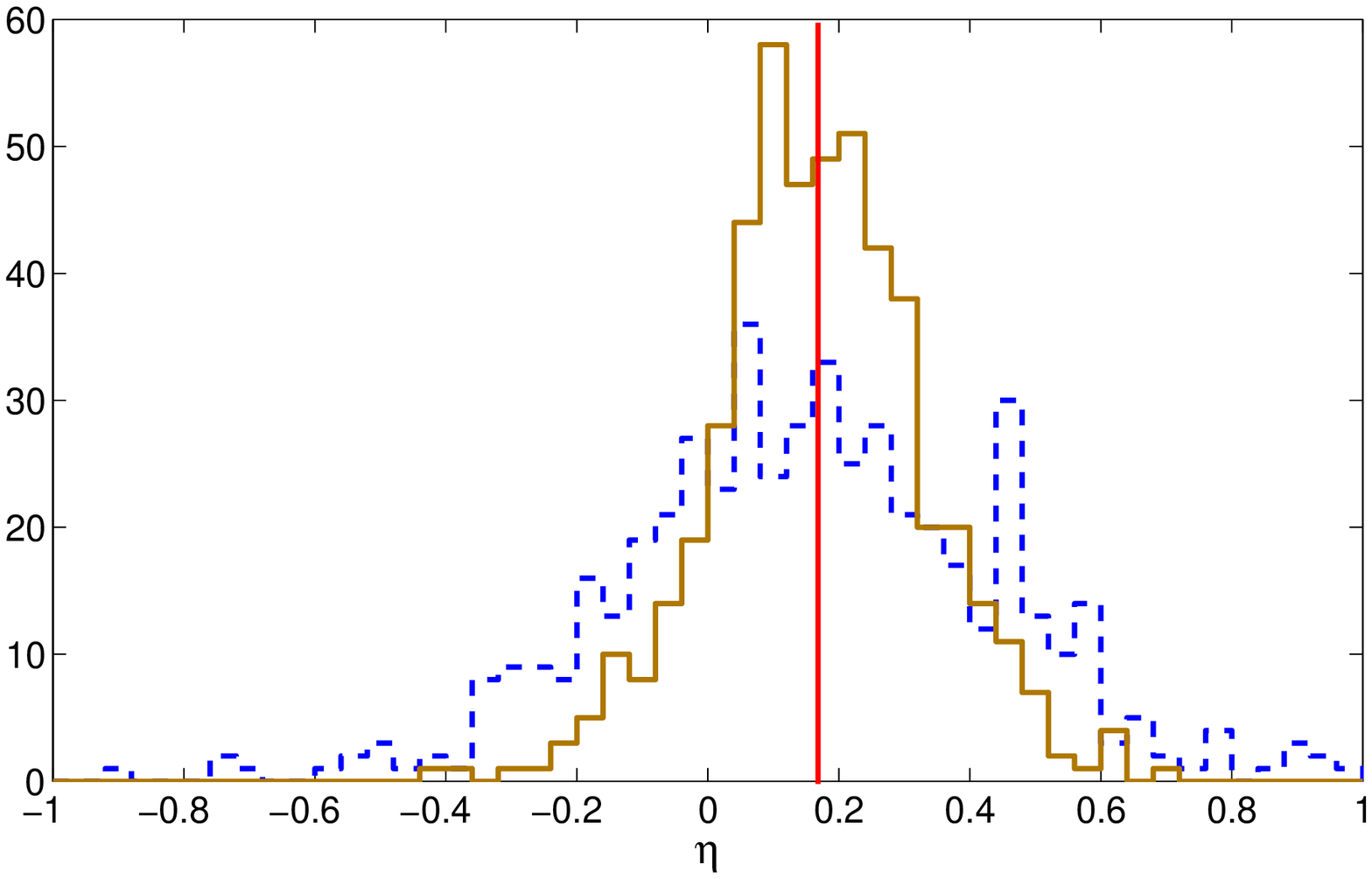}
\end{tabular}
\caption{Distribution of estimated $H_0$ and $\eta$ for single period analysis (blue, dashed line) and for the 5n-vector joint analysis (green, solid line) in the case $SNR=5$, see Tab.(\ref{tab2}). Vertical red lines indicate the injected parameters. Note that the estimation of parameters over a single period is clearly more biased and has a larger variance respect to the result of the joint analysis.}\label{fig1}
\end{figure}

In Tab.(\ref{tab2}) we show the results of the same kind of analysis for various values of SNR. The gain in jointly analyzing different datasets with the 5n-vector method is more evident for moderate or low signal-to-noise ratio, when the bias and variance of the estimations in the 5n-vector analysis are much smaller than that of the 5-vector analyses on the single datasets. For instance, for $SNR=5$ the average relative error on the amplitude $H_0$ is of about $11.8\%$ in the analysis of the single datasets, while it is of about $3.4\%$ for the 5n-vector analysis. Similarly, for $\eta$ we have a relative error of about $15\%$ in the former case and $6.2\%$ in the latter case. In Fig.(\ref{fig1}) the distribution of estimated parameters for the 5-vector analysis over a single dataset and for the 5n-vector analysis over two datasets are shown for $SNR=5$. The smaller bias and variance of the estimations with the 5n-vector method are clearly visible.

\begin{table}[h]
\caption{Parameter estimation in the analysis of simulated data. The simulation consisted in the injection of a simulated signal into 500 realizations of white Gaussian noise for different values of SNR. Injections were done into two datasets both of length 100 days: one assumed to belong to the Virgo detector, the other belonging to LIGO Livingston. The noise level was the same in the two cases. The values of injected parameters are the same as in Tab.(\ref{tab1}). In the first two tables the mean value and variance of the estimated parameters obtained with the 5-vector analysis applied separately to the two datasets are given. In the third table the results for the 5n-vector analysis over the two periods are reported (with n=2 in this case). Values for the polarization angle are not shown to keep the table simpler.}
\centering
\begin{tabular}{c c c c c}
\multicolumn{5}{c}{\rule[-3mm]{0mm}{8mm} Virgo} \\
\hline
SNR & $\bar{H}_0$ & $\sigma^2_{H_0}$ & $\bar{\eta}$ & $\sigma^2_{\eta}$  \\ [0.5ex]
\hline
5 & 1.1403 & 0.0417 & 0.1394 & 0.0451 \\
15 & 1.0127 & 0.0044 & 0.1635 & 0.0049  \\
30 & 1.0064 & 0.0012 & 0.1609 & 0.0013  \\[1ex]
\hline \\[1ex]
\end{tabular}

\begin{tabular}{c c c c c}
\multicolumn{5}{c}{\rule[-3mm]{0mm}{8mm} LIGO-L} \\
\hline
SNR & $\bar{H}_0$ & $\sigma^2_{H_0}$ & $\bar{\eta}$ & $\sigma^2_{\eta}$  \\ [0.5ex]
\hline
5 & 1.0962 & 0.0408 & 0.1344 & 0.0445 \\
15 & 1.0172 & 0.0045 & 0.1619 & 0.0048 \\
30 & 1.0082 & 0.0011 & 0.1603 & 0.0012 \\[1ex]
\hline \\[1ex]
\end{tabular}

\begin{tabular}{c c c c c}
\multicolumn{5}{c}{\rule[-3mm]{0mm}{8mm} 5n-vector} \\
\hline
SNR & $\bar{H}_0$ & $\sigma^2_{H_0}$ & $\bar{\eta}$ & $\sigma^2_{\eta}$  \\ [0.5ex]
\hline
5 & 1.0338 & 0.0226 & 0.1497 & 0.0240 \\
15 & 1.0081 & 0.0023 & 0.1624 & 0.0025 \\
30 & 1.0036 & 0.0006 & 0.1609 & 0.0007 \\[1ex]
\hline \\[1ex]
\end{tabular}

\label{tab2}
\end{table}  

\section{Further developments}
\label{conc}
The next step to test the method will be the analysis of real data from Virgo and LIGO detectors, searching for software and hardware injected simulated signals. After this validation it will be ready for the search of real signals. A priority will be the joint analysis of Virgo VSR2 and VSR4 and of LIGO S6 data targeted to the Crab pulsar, for which the sensitivities could allow to improve the best published upper limit \cite{crab_ul}. 

Another relevant issue to be addressed is the possible presence of a phase discontinuity in the signal to be detected. This could be due, for instance, to a glitch of the emitting source, like those observed in several pulsars. Glitches are sudden spin-up of a neutron star followed by a long term relaxation period. The mechanism at the base of glitches is still poorly known but the most popular models invoke quakes in the neutron star crust \cite{ruder} or the transfer of angular momentum from the superfluid to the inner crust \cite{itoh}. If this happens we can always think to split the data into two datasets corresponding respectively to times before and after the glitch. In this case, however, the direct application of the 5n-vector analysis would not work 
due to the relative phase of the signal in the datasets. If the phase jump can be accurately estimated, e.g. from radio-astronomical observations, then it is easy to properly shift the signal templates in order to correct it. If the phase shift is not known a simple 'brute force' approach could be to treat it as a free parameter and estimate it by numerically maximizing the total {\it coherence}. More sophisticated solutions need a detailed study. This issue will be discussed in detail elsewhere.

\end{document}